\documentclass[twocolumn]{aastex631}

\usepackage{amssymb}
\usepackage{soul}
\usepackage{natbib}
\usepackage{multirow}
\usepackage{graphicx}
\usepackage{tabularx}
\usepackage{verbatim}
\usepackage{color}
\usepackage{ulem}
\usepackage{subfigure}
\usepackage{booktabs,tikz}
\makeatletter
\global\let\tikz@ensure@dollar@catcode=\relax
\makeatother
\usepackage{urwchancal}
\newcommand\tikzmark[1]{\tikz[remember picture] \node (#1) {};}

\newcommand\soutm{\bgroup\markoverwith
{\textcolor{black}{\rule[0.5ex]{2pt}{0.8pt}}}\ULon}

\def\tcm{\textcolor{black}}

\submitjournal{ApJ}

\shorttitle{Formation of the ICL and BCG}
\shortauthors{Chun et al.}

\graphicspath{{./}{figures/}}

\begin{document}

\title{The formation of the brightest cluster galaxy and intracluster light in cosmological N-body simulations with the Galaxy Replacement Technique}

\correspondingauthor{Jihye Shin}
\email{jhshin@kasi.re.kr}

\author{Kyungwon Chun}
\affil{Korea Astronomy and Space Science Institute (KASI), 776 Daedeokdae-ro, Yuseong-gu, Daejeon 34055, Korea}
\author{Jihye Shin}
\affil{Korea Astronomy and Space Science Institute (KASI), 776 Daedeokdae-ro, Yuseong-gu, Daejeon 34055, Korea}
\author{Rory Smith}
\affil{Departamento de F{\'i}sica, Universidad T{\'e}cnica Federico Santa Mar{\'i}a, Avenida Vicu{\~n}a Mackenna 3939, San Joaqu{\'i}n, Santiago, Chile}
\author{Jongwan Ko}
\affil{Korea Astronomy and Space Science Institute (KASI), 776 Daedeokdae-ro, Yuseong-gu, Daejeon 34055, Korea}
\affil{University of Science and Technology (UST), Gajeong-ro, Daejeon 34113, Korea}
\author{Jaewon Yoo}
\affil{Quantum Universe Center, Korea Institute for Advanced Study (KIAS), 85 Hoegiro, Dongdaemum-gu, Seoul 02455, Korea}
\affil{Korea Astronomy and Space Science Institute (KASI), 776 Daedeokdae-ro, Yuseong-gu, Daejeon 34055, Korea}

\begin{abstract}
We investigate the formation channels of the intracluster light (ICL) and the brightest cluster galaxy (BCG) in clusters at $z=0$.
For this, we perform multi-resolution cosmological N-body simulations using the ``Galaxy Replacement Technique" (GRT).
We study the formation channels of the ICL and BCG as a function of distance from the cluster center and the dynamical state of the clusters at $z=0$. To do this, we trace back the stars of the ICL and BCG, and identify the stellar components in which they existed when they first fell into the clusters.
We find that the progenitors of the ICL and BCG in the central region of the cluster fell earlier and with a higher total mass ratio of the progenitors to the cluster compared to the outer region.
This causes a negative radial gradient in the infall time and total mass ratio of the progenitors.
Although stellar mass of the progenitors does not show the same radial gradient in all clusters,  massive galaxies ($M_{\rm{gal}} > 10^{10}~M_{\odot}~h^{-1}$) are the dominant formation channel of the ICL and BCG for all clusters, except for our most relaxed cluster.
For clusters that are dynamically more unrelaxed, we find that the progenitors of the ICL and BCG fall into their clusters more recently, and with a higher mass and mass ratio.
Furthermore, we find that the diffuse material of massive galaxies and group-mass halos that is formed by pre-processing contributes significantly to the ICL in the outer region of the unrelaxed clusters.
\end{abstract}


\section{Introduction}
Galaxy clusters are the largest gravitationally bound structures in the current concordance cosmology of Lambda cold dark matter ($\Lambda$CDM).
The clusters are hierarchically built up by merging smaller structures.
\tcm{As the accreted luminous substructures gravitationally interact with other structures in the cluster, they lose some components or can even be completely disrupted. 
Some of these components will contribute to the growth of the brightest cluster galaxy (BCG) \tcm{\citep[e.g.,][]{ostriker1977,richstone1983,delucia2007}}, the most massive central galaxy located in the deepest potential well of the cluster, whereas others spread throughout the cluster's potential \tcm{\citep[e.g.,][]{zibetti2005,conroy2007,murante2007,burke2015}}. 
These luminous components that become unbound from satellite galaxies and spread throughout the cluster’s potential are defined as the Intracluster light (ICL) \tcm{\citep[e.g.,][]{mihos2005,contini2014,montes2018}}.}
The characteristics of the ICL can give us insight into the evolution of the cluster, as well as trace the distribution of dark matter, which determines the shape of the potential well in the cluster \tcm{\citep[e.g.,][]{asensio2020,sampaio2021,yoo2022}}.

Since the ICL was first observed in the Coma cluster by \cite{zwicky1951}, observational studies have shown that the ICL is a ubiquitous component in the nearby clusters and even in clusters beyond $z=1$ \tcm{\citep[e.g.,][]{mihos2005,janowiecki2010,adami2013,ko2018,montes2018,jimenez-teja2019,chen2022}}.
Many previous studies have suggested several formation channels of the ICL, such as in-situ star formation \tcm{\citep[e.g.,][]{puchwein2010}}, tidal disruption of the dwarf galaxies \tcm{\citep[e.g.,][]{janowiecki2010}}, tidal stripping of satellite galaxies \tcm{\citep[e.g.,][]{rudick2009,puchwein2010,contini2018,montes2014,ragusa2021}}, major mergers of cluster members during the BCG formation \tcm{\citep[e.g.,][]{murante2007,contini2018}}, and pre-processing in groups \tcm{\citep[e.g.,][]{rudick2006}}.
All channels may contribute to the formation of the ICL, but the dominance may differ depending on the evolutionary stages of the clusters \tcm{\citep[e.g.,][]{ko2018}}.

Although the ICL has a low surface brightness of $\mu_{V} > 26.5~mag~arcsec^{-2}$ \citep{mihos2005,mihos2017,rudick2011}, observation and simulation studies have suggested that the ICL occupies 2\%-40\% of the total stellar component \citep{burke2015,morishita2017,jimenez-teja2018,montes2018,spavone2018,furnell2021,yoo2021}.
Observations have also shown that the ICL is extended to a few hundred kiloparsecs from the BCG, located in the central region of the cluster.
Therefore, studying the ICL and BCG combined gives us an improved understanding of the formation and evolution of clusters.

During evolution of the clusters, each has a unique mass growth history, and therefore they can be found today in a wide variety of dynamical stages.
Using 412 SDSS clusters, \cite{soares2019} found that the stellar components in dynamically unrelaxed clusters are younger than those in relaxed clusters.
This is because the relaxed clusters experience multiple merging with smaller structures earlier than unrelaxed clusters and can have more time to disrupt the accreted structures.
As the tidally disrupted objects naturally lead to the growth of the masses of the ICL and BCG, the mass fraction of the ICL and BCG in the relaxed clusters can be more significant than that in the unrelaxed clusters \tcm{\citep[e.g.,][]{rudick2006,dolag2010,contini2014,cui2014,montes2018,canas2020}}.
In addition, the different mass growth histories between the relaxed and unrelaxed clusters may imprint different stellar populations into the ICL and BCG components.

\cite{demaio2020} found that the ratio of stellar components in the inner region ($r < 10$~kpc) to the total stellar components within 100~kpc decreases as a cluster grows and suggested inside-out mass growth of the ICL and BCG within 100~kpc.
Further, \cite{contini2018} showed that BCG grows the bulk of its mass before $z=1$, whereas the diffused ICL grows its mass rapidly after $z=1$, even if the ICL and BCG coevolve after $z\sim0.7-0.8$.
This inside-out mass growth of the ICL and BCG can make a difference to the properties of the ICL and BCG depending on the distance from the cluster center.
Indeed, many previous studies have shown the negative radial profile of various properties (e.g., age, metallicity, and color) for both ICL and BCG or for the ICL alone \citep{demaio2018,montes2018,zhang2019,edwards2020,chen2022}.
For instance, \cite{montes2018} measured negative age and metallicity gradients of BCGs ($r < 50$~kpc) and ICL ($50 < r < 120$~kpc) in six clusters between $z\sim0.3$ and $z\sim0.6$.
The authors suggested that the ICL is formed by stripping Milky Way-like galaxies because the average metallicity of the ICL is comparable with the outskirts of the Milky Way.
\cite{contini2014} also showed that stripping of the massive satellites with $M_{\rm{star}} > 10^{10.5}~M_{\odot}$ are major contributors to the ICL. 

Using the Illustris-TNG300 simulation, \cite{pillepich2018} found that half of the accreted stars in both ICL and BCG come from the progenitors with $M_{\rm{star}} > 10^{11}~M_{\odot}$.
This indicates that the dominant formation channel is from massive galaxies, as suggested by \cite{contini2014} and \cite{montes2018}.
However, their results suggest that the rest of the BCG ($r < 30$~kpc) is formed by more massive progenitors than that of the ICL ($r > 100$~kpc).
\cite{murante2007} also showed that the relative importance of less massive progenitors is greater in the outer ICL ($r > 0.5R_{\rm{vir}}$) than in the central ICL ($r < 250~h^{-1}$~kpc).

The progenitors contributing to the mass growth of the ICL and BCG can grow their mass by merging with other structures and experience environmental effects before falling into the host cluster, namely pre-processing \tcm{\citep[e.g.,][]{fugita2004,mcgee2009,bianconi2018,han2018}}.
In this process, some structures suffer tidal mass loss from the tides of massive galaxies or group-sized halos and generate diffuse material similar to ICL by stripping of their main stellar body \tcm{\citep[e.g.,][]{bell2008,darocha2005,darocha2008,spavone2018}}.
Many previous studies have shown that pre-processed diffuse material contributes to the mass growth of the ICL \tcm{\citep[e.g.,][]{rudick2006,contini2014,harris2017,mihos2017}}.
In particular, \cite{harris2017} found that the ICL in the outer regions is contributed relatively greater by pre-processed diffuse material than by the main stellar body of the progenitors.

From these studies, we can predict that the formation channel of the ICL and BCG differ depending on the distance from the center as well as the dynamical state of the clusters.

In this paper, we perform multi-resolution cosmological N-body simulations using the ``Galaxy Replacement Technique" (GRT) introduced first in \cite{chun2022} to investigate the formation of the ICL and BCG in the six clusters, whose virial mass $M_{\rm{vir}}$ is $\sim 1-2\times10^{14}~M_{\odot} ~h^{-1}$ at $z=0$.
For this, we attempt to understand which progenitors contribute to the ICL and BCG and whether the properties of the progenitors are different depending on the distance from the cluster's center and/or the dynamical state of the clusters.
We also investigate the significance of pre-processing in the growth of the ICL and BCG.

This paper is organized as follows. 
In Section \ref{sec:simulation}, we introduce the GRT method and show the properties of six clusters modeled in this way.
Section \ref{sec:infall} is devoted to showing how the progenitors of the ICL and BCG differ depending on the distance from the center and the dynamical state of the clusters.
In Section \ref{sec:pre}, we investigate the significance of pre-processing on the ICL and BCG.
Lastly, we discuss and summarize our results in Section \ref{sec:discussion} and Section \ref{sec:summary}.

\section{Simulation} %%%%%%%%%%%%%%%%%%%%%%% simulation
\label{sec:simulation}

\subsection{The Galaxy Replacement Technique} %%%%%%%%%%%%%%%%%%%%%%% GRT
\label{sec:GRT}

The GRT enables us to trace out the spatial distribution and evolution of the ICL and BCG without computationally expensive baryonic physics.
This inexpensive calculation enables us to reach a higher mass and spatial resolution and to model the tidal stripping process and formation of very low surface brightness features.
To study the formation of the ICL and BCG, we first perform a low-resolution DM-only cosmological simulation of a (120 Mpc$~h^{-1}$)$^3$ uniform box with 512$^3$ particles using the cosmological simulation code Gadget-3 \citep{springel2005}.
In this simulation, the particle mass is $\sim$10$^{9}~M_{\odot}~h^{-1}$, and the gravitational softening length is fixed at 2.3 kpc$~h^{-1}$ on a comoving scale.
The initial condition for the simulation is generated at $z=200$ with the MUSIC package \footnote{https://bitbucket.org/ohahn/music/} \citep{hahn2011} using the post-Planck cosmological model of $\Omega_{m} = 0.3,~\Omega_{\Lambda} = 0.7,~\Omega_{b} = 0.047$, and $h = 0.684$.
The halos and their subhalos are identified with the modified 6D phase-space halo finder ROCKSTAR \footnote{https://bitbucket.org/pbehroozi/rockstar-galaxies} \citep{behroozi2013b} and the merger tree of these halos is built using Consistent Trees \citep{behroozi2013c}.
Among the halos at $z=0$, we select the six clusters ($M_{vir} \sim 1-2\times10^{14}~M_{\odot} ~h^{-1}$) to perform the resimulation with higher resolution particles.
We refer to these clusters as the ``GRT clusters". 
\tcm{The GRT clusters are less massive than Virgo or other clusters \citep{montes2014,montes2018,mihos2017,demaio2018} that the formation channels of the ICL have been actively studied. However, in this study, we deliberately choose this narrow mass range to investigate how the formation of the ICL and BCG depends on the dynamical state of the clusters at an approximately fixed cluster mass.}
We describe the properties of the clusters later in Section \ref{sec:cluster}.

From the merger tree, we identify all the halos that will later contribute to the mass growth of the six GRT clusters and their satellites.
To trace the evolution of the GRT clusters with higher resolution particles, we replace each low-resolution DM halo with a high-resolution model that consists of a high-resolution DM halo and a stellar disk composed of high-resolution star particles.
As the halos with $M_{\rm{peak}} < 10^{11}~M_{\odot}~h^{-1}$ consist of less than 100 low-resolution DM particles and would be expected to contribute less than 2\% of the total stellar mass in the clusters at $z=0$, we only replace halos with $M_{\rm{peak}} > 10^{11}~M_{\odot}~h^{-1}$, where $M_{\rm{peak}}$ is the maximum mass of a DM halo before falling into a more massive halo.
\tcm{
In this work, a halo is replaced with the high-resolution model when one of the following criteria is first satisfied; (1) the halo reaches $M_{\rm{peak}}$, (2) the halo first accretes a replaced satellite. 
}
In the high-resolution model, we assume an NFW density profile and a bulge-less exponential disk galaxy \footnote{\tcm{\cite{dutton2009} showed most low-mass disk galaxies with a stellar mass of $M_{\rm{star}} < 10^{10}~M_{\odot}$ have an exponential disk morphology (S{\'e}rsic index $n$ $<$ 1.5). Furthermore, \cite{shibuya2015} found that typical star-forming galaxies in this mass range have a value of $n$ $\sim$ 1-1.5 from $z=6$ to $z=0$. In contrast, for more massive galaxies, the value of $n$ is 1-1.5 until $z\sim1$ but increases after $z\sim1$ even if they are star-forming galaxies. Therefore, our assumption broadly agrees with most galaxies except for the massive galaxies replaced after $z\sim1$. Note that the fraction of these massive galaxies replaced after $z\sim1$ is only 5\% in the simulations with the six GRT clusters.}} for the DM halo and stellar disk, respectively.
The mass of the stellar disk is determined using the stellar-to-halo mass relation as a function of the redshift from \cite{behroozi2013a}. 
\tcm{The scale length of the stellar disk is determined using the mass-size relation of observed galaxies found by \cite{dutton2011}, and the scale height is fixed at 15\% of the scale length.}
We find that the ICL fractions are not strongly dependent on the choice of abundance matching models that we use (see \citealp{chun2022}). 

To complete the simulation with the GRT, we perform the multi-resolution resimulation on six clusters, and follow their evolution until $z=0$.
The gravitational softening length for the high-resolution DM and stellar particles of $\sim$ 100 and $\sim$ 10 pc$~h^{-1}$, respectively.
The high-resolution particle mass for DM and star is 5.4$\times$10$^{6}~M_{\odot}~h^{-1}$ and 5.4$\times$10$^{4}~M_{\odot}~h^{-1}$, respectively.
The softening length and mass of low-resolution DM particles are the same as those of the low-resolution DM-only simulation.
The mass of high-resolution DM (or star) particles is $\sim$200 (or 20,000) times smaller than that of the low-resolution DM particles.
\tcm{Although the presence of low-resolution DM particles could, in principle, artificially affect the evolution of the high-resolution stellar disk, the larger softening used for the low-resolution DM particles largely suppresses the scattering of high-resolution particles. Our measurements of the size evolution of infalling galaxies reveal that the level of artificial heating by low-resolution cluster DM particles is not significant for the studies of the ICL and BCG formation.}

In this work, we define the most massive galaxy as the BCG and the stellar components that are gravitationally bound to the cluster but not bound to any satellite galaxy as the ICL.
The 6D phase-space halo finder ROCKSTAR allows us to separate the BCG and ICL from the subhalos of the GRT clusters in the phase-space plane.
According to this definition, the intra-group light (IGL) of group-sized halos and stellar halo of massive galaxies in the cluster are not identified as the ICL and BCG.
We discuss the influence of their presence on the amount of the ICL in Section \ref{sec:discussion}.

\subsection{Dynamical properties of GRT clusters} %%%%%%%%%%%%%%%%%%%%%%% GRT cluster 
\label{sec:cluster}

\begin{deluxetable}{cccccc}
\tablenum{1}
\tablecaption{{The properties of the GRT clusters} \label{tab:clusters}} 
\tablewidth{0pt}
\tablehead{
 & $M_{\rm{vir}}$ & \multicolumn1c{$M_{\rm{star}}$} & $M_{\rm{gal}}$ & f$_{\rm{ICL+BCG}}$ & $z_{\rm{m50}}$ \\
 & & ($M_{\odot}~h^{-1}$) & & &
}
\startdata
C1 & $2\times10^{14}$ & $1.3\times10^{12}$ & {$1.0\times10^{12}$} & 0.180 & 0.174 \\
C2 & $2\times10^{14}$ & $1.0\times10^{12}$ & {$7.4\times10^{11}$} & 0.256 & 0.353 \\
C3 & $1.1\times10^{14}$ & $6.6\times10^{11}$ & {$3.4\times10^{11}$} & 0.478 & 0.384 \\
C4 & $1.1\times10^{14}$ & $4.6\times10^{11}$ & {$3.1\times10^{11}$} & 0.334 & 0.955 \\
C5 & $1.5\times10^{14}$ & $7.6\times10^{11}$ & {$4.0\times10^{11}$} & 0.469 & 1.106 \\
C6 & $2\times10^{14}$ & $1.0\times10^{12}$ & {$5.1\times10^{11}$} & 0.492 & 1.386 \\
\enddata
\tablecomments{Columns 2-5 are the quantities at $z=0$: virial mass, total stellar mass, surviving satellites' stellar mass, and the fraction of combined ICL and BCG, respectively. Column 6 is the redshift when each cluster first reaches half of the virial mass at $z=0$.}
\vspace{-1cm}
\end{deluxetable}

We perform the GRT simulations with six cosmological clusters of $M_{\rm{vir}} \sim 1-2\times10^{14}~M_{\odot}~h^{-1}$.
We name these GRT clusters C1-C6 (Table \ref{tab:clusters}).
The table shows that even if the clusters have similar virial mass at $z=0$, their total stellar mass ($M_{\rm{star}}$), surviving satellites' stellar mass ($M_{\rm{gal}}$), and the fraction of combined ICL and BCG (f$_{\rm{ICL+BCG}}$) vary between them.
These differences arise from the unique mass growth history of each cluster. 
To quantify the mass growth history of the clusters, we use the $z_{\rm{m50}}$ parameter, defined as the epoch when the cluster first acquires half of its virial mass at $z=0$.
Table \ref{tab:clusters} is ordered by $z_{\rm{m50}}$.

\begin{figure*}
\centering
\includegraphics[width=0.73\textwidth]{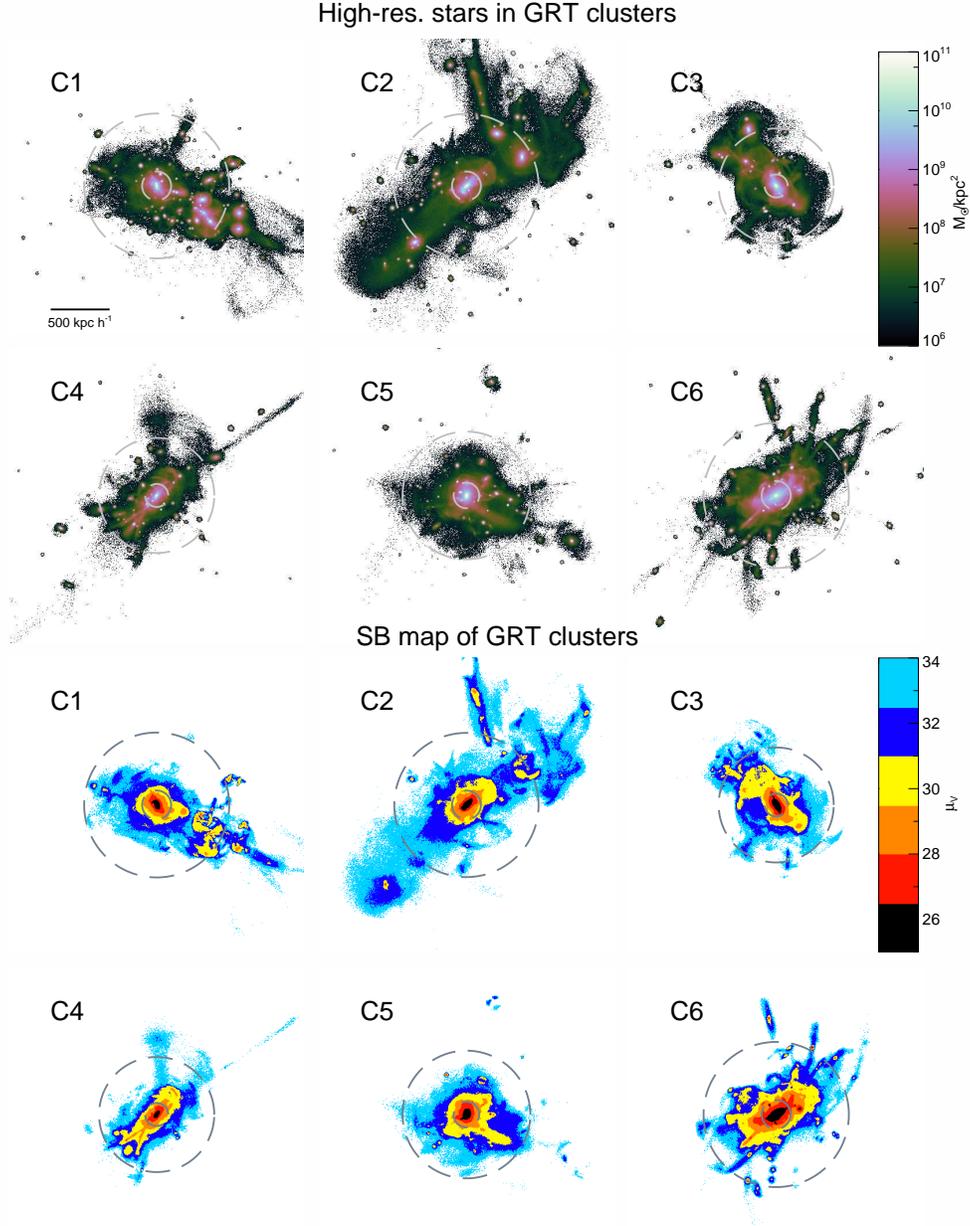}
\caption{The stellar distribution maps of the six GRT clusters at $z=0$. The upper six panels show all the stellar components colored by stellar surface density. The lower six panels show V-band surface brightness ($\mu_{V}$) of the ICL and BCG components, except for the satellites. The inner and outer gray dashed circular lines indicate the 0.1 and 0.5 $R_{\rm{vir}}$ of each cluster, respectively. Color bars indicate the surface density and V-band surface brightness.}
\label{fig:structure}
\end{figure*}

As each cluster has a unique mass growth history, the stellar structures of the six GRT clusters differ greatly.
Figure \ref{fig:structure} shows the stellar structures of the six GRT clusters at $z=0$. The upper six panels show all stellar structures, including the ICL, BCG, and satellites, and the lower six panels show the V-band surface brightness ($\mu_{V}$) of the ICL and BCG inside the six GRT clusters. \tcm{Because we cannot know the exact age or metallicity of each stellar particle in the GRT simulations, we convert from the stellar mass distribution to the V-band surface brightness map using a constant and typical V-band mass-to-light ratio of 5$M_{\odot}/L_{\odot}$ \citep{bruzual2003,longhetti2009,burke2012}.} \tcm{In this figure,} we use a V-band surface brightness limit of $\mu_{V}=26.5~mag~arcsec^{-2}$ to separate the galaxy from the ICL for easier \footnote{\tcm{Note that this separation is only used in Figure \ref{fig:structure}.}}. Following this, the central black-colored region would be defined as the BCG. The inner and outer gray dashed lines indicate the $0.1R_{\rm{vir}}$ and $0.5R_{\rm{vir}}$, respectively.
This indicates that most of the BCG components are located within $0.1R_{\rm{vir}}$.
\tcm{In the surface brightness map, we can see some compact ICL components (e.g., black point-like structures in C6). We find that these are indeed bound only to the host cluster, not the satellite, since most of the DM halo of the satellite is already stripped.}

In Figure \ref{fig:structure}, we can see that C1, which has the smallest $z_{\rm{m50}}$, experiences an ongoing major merger and has many surviving satellites at $z=0$.
Thus, C1 looks like a dynamically unrelaxed system. 
On the other hand, C6, which has the highest $z_{\rm{m50}}$, experiences the last major merger at $z=1.7$ and grows its mass slowly after then.
This makes the stellar structure of C6 more relaxed than others.
Therefore, the $z_{\rm{m50}}$ parameter seems to represent the dynamical state of the GRT clusters quite well.
Indeed, previous studies have shown the existence of a relation between $z_{\rm{m50}}$ and the dynamical state of the clusters in cosmological simulations \citep{power2012,mostoghiu2019,haggar2020}. 

In this work, we classify the six GRT clusters into two groups, unrelaxed and relaxed clusters, to study the formation channels of the ICL and BCG according to their $z_{\rm{m50}}$ value. 
I.e., the three clusters with lower $z_{\rm{m50}}$, C1-C3, are classified as the unrelaxed clusters and the others, C4-C6, as the relaxed clusters.
Note that we only divide the clusters in half according to the $z_{\rm{m50}} $, and this choice is not as robust as the criteria used in some previous studies \tcm{\citep{neto2007,power2012,cui2017}}.
\tcm{However, in the case of the six GRT clusters, we find that the $z_{\rm{m50}} $ is sufficient to substitute the parameters used in the previous studies (\citealp[virial ratio, substructure mass fraction, and center-of-mass offset;][]{neto2007,cui2017} and \citealp[virial ratio;][]{power2012}). See their paper for detailed criteria.}
We find that the unrelaxed clusters experienced the last major merger more recently than the relaxed clusters.

\begin{figure*}
\centering
\includegraphics[width=0.95\textwidth]{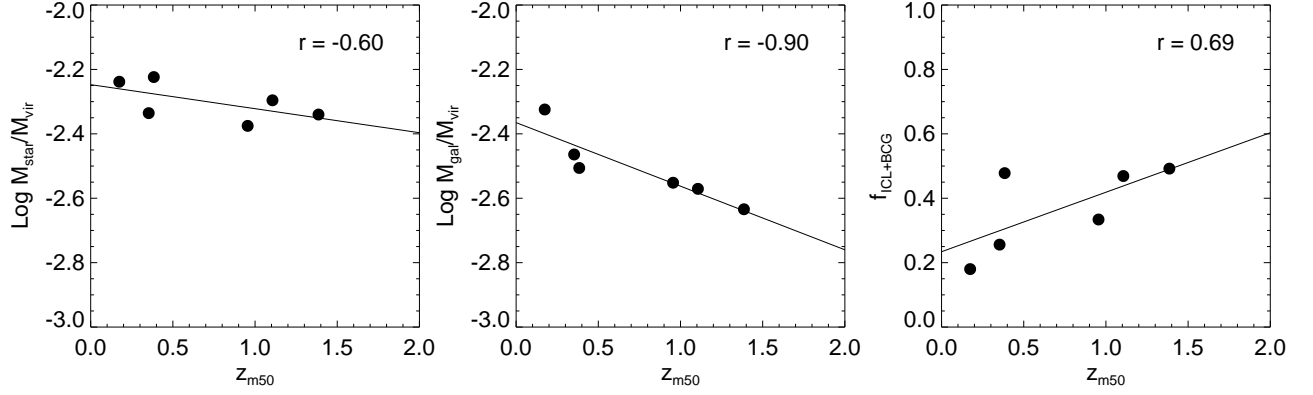}
\caption{The relations between $z_{\rm{m50}}$ and the properties at $z=0$ of the GRT clusters. From left to right, each panel shows the ratio of the total stellar mass to the virial mass, $M_{\rm{star}}$/$M_{\rm{vir}}$, the ratio of the surviving satellites' stellar mass to the virial mass, $M_{\rm{gal}}$/$M_{\rm{vir}}$, and the fraction of combined ICL and BCG, $f_{\rm{ICL+BCG}}$, as a function of $z_{\rm{m50}}$, respectively. In each panel, the filled black circles indicate the six GRT clusters. The solid black line is the best fitting line, and the value in the upper right corner indicates the Pearson correlation coefficient.} 
\label{fig:relation}
\end{figure*}

The different dynamical states of the GRT clusters result in not only different stellar structures, but also different quantitative properties of stellar components in the clusters.
Figure \ref{fig:relation} shows the relation between the stellar properties and the $z_{\rm{m50}}$ of the six GRT clusters.
The left and middle panels show that the ratio of the total stellar mass ($M_{\rm{star}}$) and surviving satellites' stellar mass ($M_{\rm{gal}}$) to the virial mass ($M_{\rm{vir}}$) of each cluster decreases with the increase of $z_{\rm{m50}}$. 
Especially, $M_{\rm{gal}}$/$M_{\rm{vir}}$ decreases most rapidly with increasing $z_{\rm{m50}}$.
This means that more galaxies were disrupted after falling into the clusters when they had higher $z_{\rm{m50}}$ because those that fall in earlier have more time to be disrupted within the clusters.
These disrupted stellar components naturally contribute to the mass growth of the ICL and BCG, and thus f$_{\rm{ICL+BCG}}$ is proportional to $z_{\rm{m50}}$ (the right panel of Figure \ref{fig:relation}).

\begin{figure}[hbp]
\centering
\includegraphics[width=0.45\textwidth]{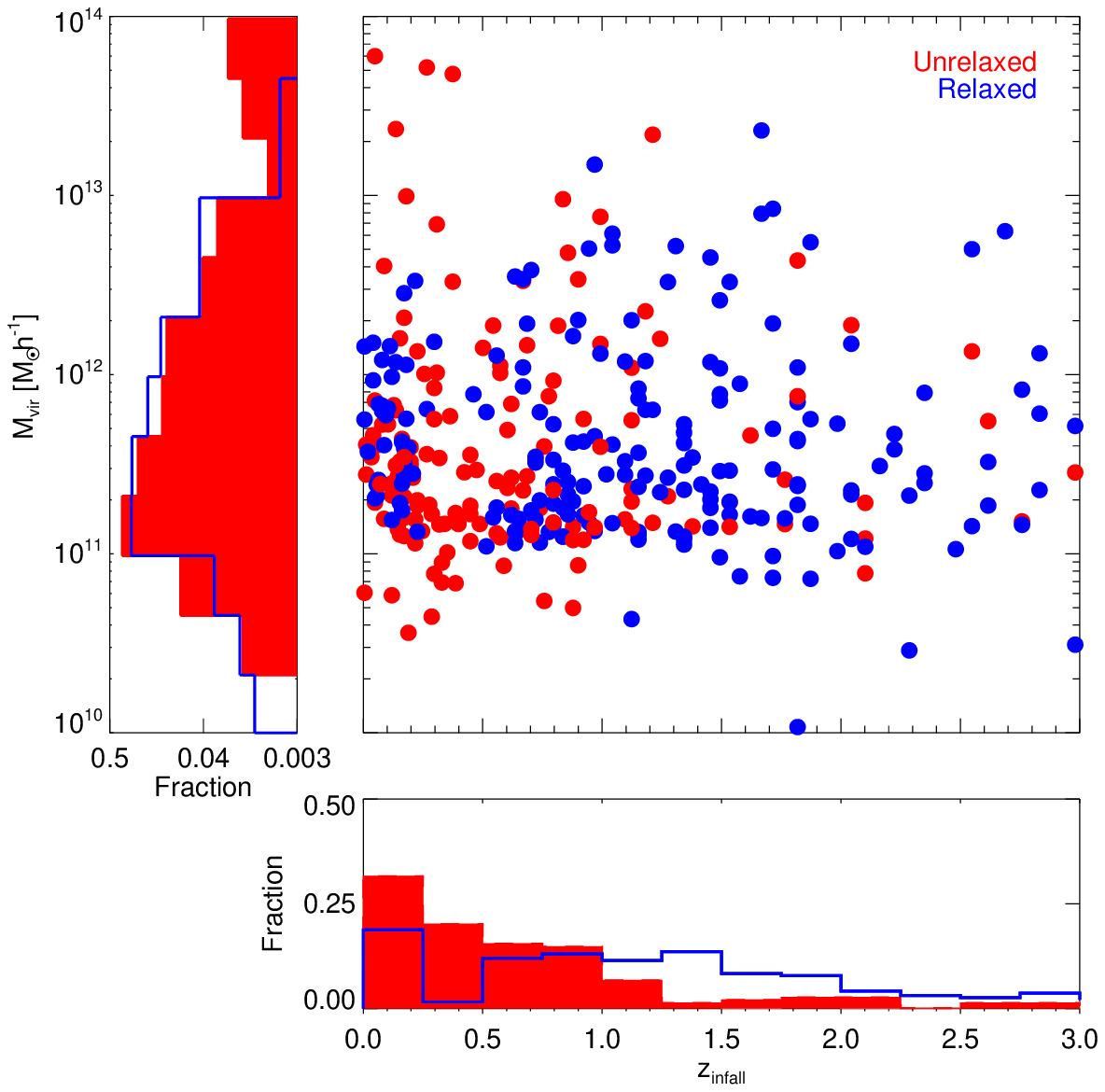}
\caption{Mass of \tcm{luminous} satellites when they first fall into their clusters. A scatter plot indicates the infall time and mass distribution of the satellites falling into the clusters. The lower histogram shows the redshift when the satellites enter the virial mass of clusters, $z_{\rm{infall}}$. The left histogram indicates the mass function of satellites at $z_{\rm{infall}}$. In the scatter plot and histograms, the relaxed and unrelaxed clusters are colored red and blue, respectively.}
\label{fig:host}
\end{figure}

Figure \ref{fig:host} shows the infall redshift $z_{infall}$ and mass distribution of the satellites falling into the relaxed and unrelaxed clusters\tcm{, regardless of the survival of satellites until $z=0$}. The infall redshift is defined as the moment when the satellites first cross the virial radius of the host cluster.
Here, we show only the host structures\tcm{, including stars} at $z_{\rm{infall}}$, even if they have substructures, and $M_{\rm{vir}}$ of the host structure includes the mass of substructures.

As the infalling satellites experience multiple merging with other halos before falling into the cluster, their mass can increase if they fall into the cluster more recently.
\tcm{However, a few satellites lose their mass by interaction with other halos before falling into the cluster. This is why there are infalling satellites of $M_{\rm{vir}} < 10^{11}~M_{\odot}~h^{-1}$ at $z_{\rm{infall}}$ although we only replace the low-resolution DM halos of $M_{\rm{vir}} > 10^{11}~M_{\odot}~h^{-1}$ with the high-resolution model.}

The lower histogram shows the fraction of satellites depending on the $z_{\rm{infall}}$.
In this histogram, the satellites falling into the unrelaxed clusters (C1-C3; filled red histograms) tend to have a lower redshift of infall than the relaxed clusters (C4-C6; blue histograms).

Indeed, the median of $z_{\rm{infall}}$ of satellites falling into the unrelaxed and relaxed clusters is 0.5 and 1.2, respectively.

The left histogram shows the mass function of infalling satellites.
The mass function of all clusters is similar, but the unrelaxed clusters show an extended tail to high masses in the mass function due to the presence of the massive groups ($M_{\rm{vir}} > 10^{13}~M_{\odot}~h^{-1}$) that fall into the clusters after $z=1$.

These two histograms show that the satellites that fall into the unrelaxed or relaxed clusters have different properties at $z_{\rm{infall}}$. 
As the ICL and BCG are formed by the disruption of the satellites, we can expect that the different channels can dominate the formation of the ICL and BCG depending on the dynamical state of the clusters.

\section{Formation channels of the ICL and BCG} %%%%%%%%%%%%%%%%infall
\label{sec:infall}
Clusters are the most massive structures that hierarchically form by merging smaller structures.
Therefore, it is natural that the ICL and BCG form by the disruption of these smaller structures.
Indeed, many studies have shown their contribution to the formation of the ICL and BCG, but the authors did not reach an agreement about the dominant formation channel \citep{contini2014,morishita2017,demaio2018,montes2018,pillepich2018}.

In this section, to better understand the formation channels of the ICL and BCG \footnote{\tcm{As the GRT does not describe the in-situ star formation in the BCG, we miss the contribution of the in-situ star formation to the growth of ICL and BCG. Due to this limitation, we only focus on the external origin of the ICL and BCG growth. However, note that most stars in the ICL and BCG are still expected to be brought in by accreted progenitors \citep[e.g.,][]{lee2017,pillepich2018,davison2020}}}, we investigate the properties at the infall time of progenitors that will contribute to the ICL and BCG and show how their properties are related to the dynamical state of the cluster.

Furthermore, as the properties of the ICL and BCG show radial gradients \tcm{\citep[e.g.,][]{montes2014,montes2018,contini2018,demaio2018}}, we can expect that the dominant formation channel also differs depending on the distance from the center of the cluster.

To investigate this, we focus on the formation channels of the ICL components in the outer region of the cluster ($R > 0.5R_{\rm{vir}}$; Section \ref{sec:rvir1}) and the ICL and BCG within the inner regions ($R < 0.1R_{\rm{vir}}$; Section \ref{sec:rvir2}), considering the two regions separately.
We name the ICL components in the outer region and the ICL and BCG within an inner region as the `outer ICL' and the `central ICL+BCG', respectively.
In Section \ref{sec:profile}, we investigate the radial gradient of the properties of the progenitors contributing to the ICL and BCG in the clusters.

\subsection{Formation channels of the ICL in outer $0.5R_{\rm{vir}}$} %%%%%%%%%%%%%%%%0.5Rvir
\label{sec:rvir1}

\begin{deluxetable*}{rlcccccccccccc}[htp]
\tablenum{2}
\tablecaption{Contribution to the `outer ICL' depending on properties of the progenitors \label{tab:radius05}} 
\tablewidth{0pt}
\tablehead{
\colhead{} & \colhead{} & \multicolumn3c{Infall time\tablenotemark{*}} & \colhead{} & \multicolumn3c{Galaxy mass\tablenotemark{**}}  & \colhead{} & \multicolumn3c{Mass ratio\tablenotemark{***}} \\
\cline{3-5} \cline{7-9} \cline{11-13}
\colhead{ } & \colhead{ } & \colhead{$\tau_1$} & \colhead{$\tau_2$} & \colhead{$\tau_3$} & \colhead{} & \colhead{$\mathcal{M}_1$} & \colhead{$\mathcal{M}_2$} & \colhead{$\mathcal{M}_3$} & \colhead{} & \colhead{$\gamma_1$} & \colhead{$\gamma_2$} & \colhead{$\gamma_3$} & \colhead{}
}
\startdata
& \tikzmark{a}{C1} & \textbf{1.00} & 0.00 & 0.00 & & 0.15 & 0.15 & \textbf{0.70} & & \textbf{0.68} & 0.04 & 0.28\\
\multirow{4}{0em}{\rotatebox{90}{$z_{\rm{m50}}$}}& \tikzmark{c}{C2} & \textbf{0.95} & 0.03 & 0.02 & & 0.13 & 0.17 & \textbf{0.65} & & 0.33 & \textbf{0.56} & 0.11\\
& \tikzmark{d}{C3} & \textbf{0.84} & 0.12 & 0.04 & & 0.18 & \textbf{0.65} & 0.17 & & \textbf{0.83} & 0.06 & 0.12\\
& \tikzmark{e}{C4} & \textbf{0.81} & 0.11 & 0.09 & & 0.19 & \textbf{0.81} & 0.00 & & \textbf{0.76} & 0.24 & 0.00\\
& \tikzmark{f}{C5} & 0.48 & \textbf{0.49} & 0.02 & & 0.36 & \textbf{0.64} & 0.00 & & \textbf{0.91} & 0.09 & 0.00\\
& \tikzmark{b}{C6} & 0.20 & \textbf{0.72} & 0.08 & & \textbf{0.63} & 0.10 & 0.27 & & \textbf{0.72} & 0.27 & 0.00\\
\enddata
\tikz[remember picture,overlay] \draw[->] (a.center -| b.center) -- (b.center);
\tablecomments{Contribution of the progenitors to the `outer ICL' depending on their infall time, galaxy stellar mass, and total mass ratio of the progenitors to cluster. The dominant formation channel is indicated in bold.}
\tablenotetext{}{*$\tau_1$, $\tau_2$, and $\tau_3$ indicate `recent infaller' ($t_{\rm{infall}}$ [Gyr] $< 8$), `intermediate-time infaller' (8$< t_{\rm{infall}}$ [Gyr] $<$~10), and `early infaller' ($t_{\rm{infall}}$ [Gyr] $ > 10$).}
\tablenotetext{}{**$\mathcal{M}_1$, $\mathcal{M}_2$, and $\mathcal{M}_3$ indicate the `low-mass infaller' ($M_{\rm{gal}} [M_{\odot}~h^{-1}] < 10^{10}$), `intermediate-mass infaller' ($10^{10} < M_{\rm{gal}} [M_{\odot}~h^{-1}] < 10^{11}$), and `high-mass infaller' ($M_{\rm{gal}} [M_{\odot}~h^{-1}] > 10^{11}$).}
\tablenotetext{}{***$\gamma_1$, $\gamma_2$, and $\gamma_3$ indicate the `low-ratio infaller' ($M_{\rm{s}}/M_{\rm{h}} < 0.1$), `minor merger infaller' ($0.1 < M_{\rm{s}}/M_{\rm{h}} < 0.33$), and `major merger infaller' ($M_{\rm{s}}/M_{\rm{h}} > 0.33$).}
\end{deluxetable*}

\begin{figure*}
\centering
\includegraphics[width=0.9\textwidth]{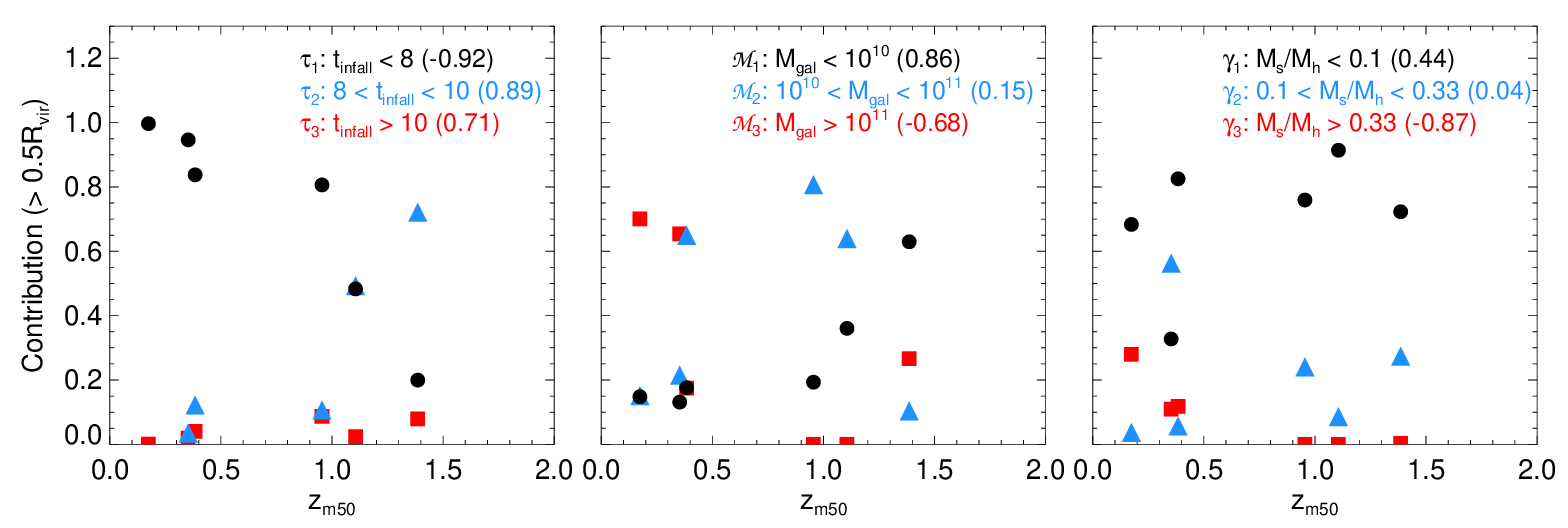}
\caption{The relation between the $z_{\rm{m50}}$ of clusters and the contribution of progenitors of the `outer ICL' according to the infall time ($t_{\rm{infall}}$), galaxy stellar mass ($M_{\rm{gal}}$), and total mass ratio of the ICL progenitor to the cluster ($M_{\rm{s}}/M_{\rm{h}}$). $M_{\rm{gal}}$ and $M_{\rm{s}}/M_{\rm{h}}$ are the properties at $t_{\rm{infall}}$. The filled black circles, blue triangles, and red squares indicate the three subgroups depending on the properties of progenitors. The meaning of each subgroup ($\tau_x$, $\mathcal{M}_x$, and $\gamma_x$) is described in Section \ref{sec:rvir1}. The values in the upper right corner are the Pearson correlation coefficients.} 
\label{fig:cont05}
\end{figure*}

As the BCG is concentrated in the central region of the cluster, recent studies have suggested that the ICL dominates in the region of $R > 20\sim100~\rm{kpc}$ \citep{gonzalez2021,montes2021,contini2022}.
Furthermore, \cite{chen2022} measured the light from the BCG and ICL of $\sim$3000 clusters between $0.2 < z < 0.3$ in the SDSS and confirmed that the ICL predominates beyond 70$\sim$200~kpc.

\tcm{The distinction between the ICL and BCG is not unique in the literature, and it might be arbitrary to define the radius outside which the ICL dominates over the BCG, as discussed in \cite{pillepich2018}.
To significantly reduce the effect of the BCG and focus on the formation channel of the ICL, we investigate the `outer ICL' components in the region of $R > 0.5R_{\rm{vir}}$ (716$\sim$906 kpc)}.
Even if it is a more outer region than the previous studies have suggested, \cite{murante2007} found that galaxies with a range of masses contribute to the ICL components in this outer region only. \tcm{Furthermore, the contribution of the BCG to the ICL is significantly reduced in this region \citep[e.g.,][]{murante2007,harris2017}.} Therefore, it is interesting to see how the properties of the infalling galaxies contributing to the ICL in this outer region depend on the dynamical state of the clusters.

In this region of the six GRT clusters, the amount of `outer ICL' is relatively small, corresponding to only 1-2\% of the total stellar mass in ICL and BCG within $R_{\rm{vir}}$ except for in C2, where it reaches $\sim$9\%, due to the \tcm{stars stripped from massive groups}.
This is discussed further in Section \ref{sec:discussion}.
However, the mass resolution of stellar particles in the GRT simulations is high ($5.4\times10^{4}~M_{\odot}~h^{-1}$) enough to resolve the `outer ICL' using $\sim10^5$ particles.

Table \ref{tab:radius05} shows the contribution of progenitors to the `outer ICL' depending on their infall time, galaxy stellar mass, and total mass ratio of the progenitors to cluster. 
Each contribution is normalized by the total stellar mass of the `outer ICL' of each cluster.
To investigate the dominant formation channel, we categorize the progenitors into three subgroups depending on the following properties of progenitors:

\begin{enumerate}
\item[(1)] \textbf{Infall time ($t_{\rm{infall}}$)}. The progenitors are categorized as the `recent infaller' ($\tau_1$; $t_{\rm{infall}} < 8$~Gyr), `intermediate-time infaller' ($\tau_2$; 8~Gyr~$< t_{\rm{infall}} <$~10~Gyr), and `early infaller' ($\tau_3$; $t_{\rm{infall}} > 10$~Gyr). The lookback times of 8 and 10~Gyr correspond to $z\sim$1 and 1.7, respectively.
\item[(2)] \textbf{Galaxy stellar mass ($M_{\rm{gal}}$)}. The progenitors are categorized as the `low-mass infaller ($\mathcal{M}_1$; $M_{\rm{gal}} < 10^{10}~M_{\odot}~h^{-1}$), `intermediate-mass infaller' ($\mathcal{M}_2$; $10^{10}~M_{\odot}~h^{-1} < M_{\rm{gal}} < 10^{11}~M_{\odot}~h^{-1}$), and `high-mass infaller' ($\mathcal{M}_3$; $M_{\rm{gal}} > 10^{11}~M_{\odot}~h^{-1}$).
\item[(3)] \textbf{Mass ratio ($M_{\rm{s}}/M_{\rm{h}}$)}. The progenitors are categorized as the 'low-ratio infaller' ($\gamma_1$; $M_{\rm{s}}/M_{\rm{h}} < 0.1$), `minor merger infaller' ($\gamma_2$; $0.1 < M_{\rm{s}}/M_{\rm{h}} < 0.33$), and `major merger infaller' ($\gamma_3$; $M_{\rm{s}}/M_{\rm{h}} > 0.33$).
\end{enumerate}

$M_{\rm{gal}}$ and $M_{\rm{s}}/M_{\rm{h}}$ of progenitors are measured at their infall time.
We arrange Table \ref{tab:radius05} in order of $z_{\rm{m50}}$ and highlight the dominant formation channel in bold.

$\tau_1$-$\tau_3$ in Table \ref{tab:radius05} show that $t_{\rm{infall}}$ of the progenitors of the `outer ICL' is related to the $z_{\rm{m50}}$ of the clusters.
The `outer ICL' in C1 and C2 is generated by the `recent infallers' in general, while the `intermediate-time infallers' are the dominant formation channel of `outer ICL' in C5 and C6.
In particular, the contribution of `recent infallers' is strongly inverse proportional to the $z_{\rm{m50}}$ of the clusters.

The left panel of Figure \ref{fig:cont05} shows more directly how the relation between the $z_{\rm{m50}}$ of clusters and the contribution of progenitors of the `outer ICL' depends on $t_{\rm{infall}}$.
The values in the upper right corner are the Pearson correlation coefficients of each infaller.
This panel shows that the contribution of the `recent infallers' (the filled black circles) decreases with increasing $z_{\rm{m50}}$ of the clusters.
Indeed, the Pearson correlation coefficient of the `recent infallers' is larger than that of the `intermediate-time infallers' and `early infallers' (the filled blue triangles and red squares).

\tcm{As the `recent infallers' have more time before falling into the cluster, they naturally have a longer period to grow their stellar mass outside the cluster. Therefore, more `recent infallers' in the unrelaxed clusters allow more massive progenitors of `outer ICL' than those in the relaxed clusters.
Indeed, }$\mathcal{M}_1$-$\mathcal{M}_3$ in Table \ref{tab:radius05} shows that the dominant formation channel of the `outer ICL' is less massive as the $z_{\rm{m50}}$ of clusters increases.
The `outer ICL' in C1 and C2 generally consists of the `high-mass infallers', whereas the majority of the `outer ICL' components in the most relaxed cluster, C6, form from `low-mass infallers'.
In addition, the contribution of the `low-mass infallers' increases with the $z_{\rm{m50}}$ of clusters. 
This trend is clearly visible in the middle panel of Figure \ref{fig:cont05} (see the filled black circles).

\cite{pillepich2018} showed that only 10\% of ICL outside 100~kpc of the clusters comes from galaxies less massive than $10^{9.8}~M_{\odot}$ and this contribution is independent of the halo mass according to the Illustris-TNG 300 simulation.
In our results, the `outer ICL' of C1-C4 also shows that the `low-mass infallers' do not contribute significantly.
On the other hand, the `low-mass infallers' can contribute significantly to the `outer ICL' of C5 and C6.
This is visible in the middle panel of Figure \ref{fig:cont05} (see the filled black circles).
However, because we investigate the ICL components in the more distant outer regions than \cite{pillepich2018}, it is hard to compare directly.
Nevertheless, our results show that the `low-mass progenitors' with $M_{\rm{gal}} < 10^{10}~M_{\odot}~h^{-1}$ can contribute significantly to the `outer ICL', and this contribution depends on the dynamical state of the clusters.

In contrast, the $M_{\rm{s}}/M_{\rm{h}}$ of the progenitors of the `outer ICL' seems to be unrelated with the dynamical state of the clusters ($\gamma_1$-$\gamma_3$ in Table \ref{tab:radius05} and the right panel of Figure \ref{fig:cont05}). 
The `outer ICL' of all the clusters generally comes from the `low-ratio infallers' (except C2 due to the significant contribution of the massive groups).
Interestingly, there is no contribution from `major merger infallers' to the `outer ICL' in the relaxed clusters, C4-C6. In fact, C4 and C5 do not experience any major mergers during the growth of the ICL and BCG, and the major merger that occurs in C6 only contributes to the mass growth of the inner regions ($R < 0.5R_{\rm{vir}}$).

\tcm{As the small amount of ICL stars is extended in the outer region, the observation focusing on them is very challenging, and thus the study of the `outer ICL' formation in the individual clusters is still in the realm of simulations.}

\subsection{Formation channels of the ICL and BCG within $0.1R_{\rm{vir}}$} %%%%%%%%%%%%%%%%0.1Rvir
\label{sec:rvir2}

\begin{deluxetable*}{rlcccccccccccc}[htp]
\tablenum{3}
\tablecaption{Contribution to the `central ICL+BCG' depending on the properties of the progenitors \label{tab:radius01}} 
\tablewidth{0pt}
\tablehead{
\colhead{} & \colhead{} & \multicolumn3c{Infall time [Gyr]} & \colhead{} & \multicolumn3c{Galaxy mass [$M_{\odot}~h^{-1}$]}  & \colhead{} & \multicolumn3c{Mass ratio} \\
\cline{3-5} \cline{7-9} \cline{11-13}
\colhead{ } & \colhead{ } & \colhead{$\tau_1$} & \colhead{$\tau_2$} & \colhead{$\tau_3$} & \colhead{} & \colhead{$\mathcal{M}_1$} & \colhead{$\mathcal{M}_2$} & \colhead{$\mathcal{M}_3$} & \colhead{} & \colhead{$\gamma_1$} & \colhead{$\gamma_2$} & \colhead{$\gamma_3$} & \colhead{}
}
\startdata
& \tikzmark{a}{C1} & \textbf{0.78} & 0.07 & 0.15 & & 0.06 & \textbf{0.94} & 0.00 & & 0.05 & \textbf{0.72} & 0.23\\
\multirow{4}{0em}{\rotatebox{90}{$z_{\rm{m50}}$}}& \tikzmark{c}{C2} & \textbf{0.78} & 0.07 & 0.15 & & 0.07 & 0.16 & \textbf{0.77} & & 0.08 & 0.04 & \textbf{0.88}\\
& \tikzmark{d}{C3} & 0.01 & \textbf{0.77} & 0.22 & & 0.02 & 0.23 & \textbf{0.75} & & 0.03 & 0.00 & \textbf{0.97}\\
& \tikzmark{e}{C4} & \textbf{0.54} & 0.28 & 0.19 & & 0.24 & \textbf{0.76} & 0.00 & & 0.44 & \textbf{0.56} & 0.00\\
& \tikzmark{f}{C5} & 0.01 & \textbf{0.73} & 0.26 & & 0.09 & \textbf{0.91} & 0.00 & & 0.45 & \textbf{0.55} & 0.00\\
& \tikzmark{b}{C6} & 0.00 & 0.39 & \textbf{0.61} & & 0.13 & \textbf{0.72} & 0.15 & & 0.39 & \textbf{0.49} & 0.12\\
\enddata
\tikz[remember picture,overlay] \draw[->] (a.center -| b.center) -- (b.center);
\tablecomments{Contribution of the progenitors to the `central ICL+BCG' depending on their infall time, galaxy stellar mass, and total mass ratio of the progenitors to cluster. The dominant formation channel is indicated in bold. The meaning of each subgroup ($\tau_x$, $\mathcal{M}_x$, and $\gamma_x$) is described in Section \ref{sec:rvir1}.}
\end{deluxetable*}

\begin{figure*}
\centering
\includegraphics[width=0.9\textwidth]{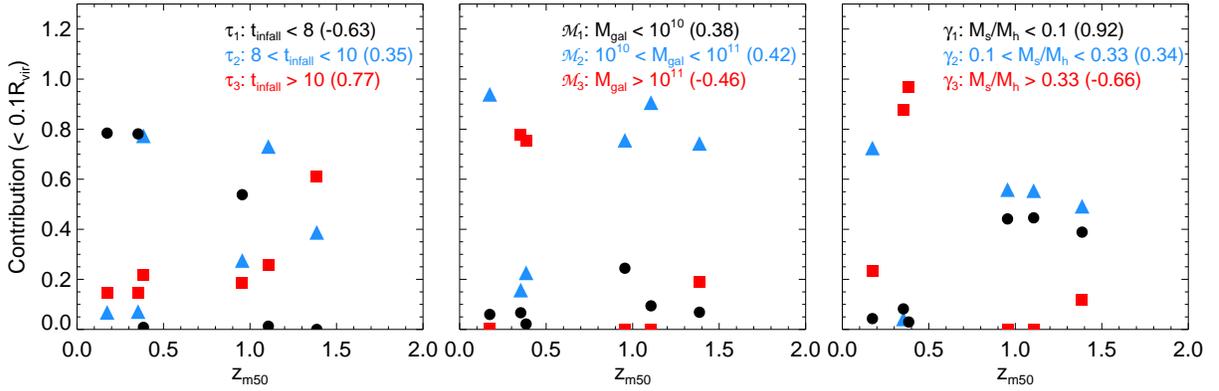}
\caption{The relation between the $z_{\rm{m50}}$ of clusters and the contribution of progenitors of the `central ICL+BCG' according to $t_{\rm{infall}}$, $M_{\rm{gal}}$, and $M_{\rm{s}}/M_{\rm{h}}$. The filled black circles, blue triangles, and red squares indicate the three subgroups depending on the properties of progenitors. The meaning of each subgroup ($\tau_x$, $\mathcal{M}_x$, and $\gamma_x$) is described in Section \ref{sec:rvir1}. The values in the upper right corner mean the Pearson correlation coefficient.}
\label{fig:cont01}
\end{figure*}

In this section, we focus on the formation of the `central ICL+BCG' of each cluster at $z=0$.
We find that most of the BCG components are located within $0.1R_{\rm{vir}}$ in the lower six panels of Figure \ref{fig:structure}.
Therefore, the choice of $0.1R_{\rm{vir}}$ (\tcm{143$\sim$181 kpc}) is reasonable for studying the BCG components as well as the ICL.
\tcm{Furthermore, this region is comparable to the region investigated for the formation channels or properties of the ICL (and BCG) in observational studies \citep[$100\sim200~$kpc, e.g.,][]{demaio2015,demaio2018,montes2018,montes2021}.}
The contents of Table \ref{tab:radius01} are the same as Table \ref{tab:radius05} in Section \ref{sec:rvir1}, but this time for the `central ICL+BCG'.

$\tau_1$-$\tau_3$ in Table \ref{tab:radius01} shows that $t_{infall}$ of the progenitors of the `central ICL+BCG' is related to the $z_{\rm{m50}}$ of the clusters.
The `central ICL+BCG' in C1 and C2 are generated by the `recent infallers' in general, whereas, those in C5 and C6 come from the progenitors that fall into the cluster earlier, $t_{\rm{infall}} > $8~Gyr.
This causes the contribution of the `early infallers' ($t_{\rm{infall}} > 10$~Gyr) to increase with $z_{\rm{m50}}$ of the clusters.
This increase is also shown in the left panel of Figure \ref{fig:cont01} (see the filled red squares).
Indeed, the Pearson correlation coefficient of the `early infallers' (the filled red squares) is larger than that of the `recent infallers' and `intermediate-time infallers' (the filled black circles and blue triangles).

The `intermediate-time infallers' and `early infallers' significantly contribute to the `outer ICL' only in clusters C5 and C6, but their contribution to the `central ICL+BCG' is significant for all clusters.
Although $t_{\rm{infall}}$ does not directly represent the formation time of the ICL and BCG components, it suggests the possibility that the `central ICL+BCG' components form earlier than the `outer ICL' components.

As we mentioned in Section \ref{sec:rvir1}, an increase in the contribution of `recent infallers' indicates that more massive progenitors at $t_{\rm{infall}}$ contribute to the mass growth of the `central ICL+BCG'.
$\mathcal{M}_1$-$\mathcal{M}_3$ in Table \ref{tab:radius01} shows it as the progenitors of the `central ICL+BCG' generally become more massive in unrelaxed clusters (with the exception of C1) than in relaxed clusters.
In the case of C1, there is no contribution from `high-mass infallers' to the `central ICL+BCG', even though this cluster experiences an ongoing merger with two group-sized halos with $M_{\rm{gal}} > 10^{11}~M_{\odot}~h^{-1}$.
We find that these group-sized halos are first infallers which have not reached their first pericenter by $z=0$.
This unfinished merging of group-sized halos in C1 causes the relation between $z_{\rm{m50}}$ and each subgroup to be unclear (the middle panel of Figure \ref{fig:cont01}).

In Section \ref{sec:rvir1}, $\gamma_1$-$\gamma_3$ in Table \ref{tab:radius05} show that most components of the `outer ICL' come from the `low-ratio infallers'.
On the other hand, the contribution from `minor merger infallers' and `major merger infallers' is more significant for the `central ICL+BCG' in all clusters (Table \ref{tab:radius01}).
In relaxed clusters, the dominant formation channel of the `central ICL+BCG' is from `minor merger infallers'.
The `central ICL+BCG' components of the most unrelaxed cluster, C1, also form from the `minor merger infallers', but this is once again because they do not reach the central region by $z=0$, as mentioned above.
If the merging event would be finished, the contribution from the `major merger infallers' would be increased, like in C2 and C3.
Even if the contribution of the progenitors with $M_{\rm{s}}/M_{\rm{h}}$ larger than 0.1 is dominant for all clusters, the `low-ratio infallers' still contribute to the `central ICL+BCG' in relaxed clusters significantly.
This significant contribution is also shown in the right panel of Figure \ref{fig:cont01} (see the filled black circles).

\subsection{Radial profile of the properties of the ICL and BCG progenitors} %%%%%%%%%%%%%%%%profile
\label{sec:profile}

In Section \ref{sec:rvir1} and \ref{sec:rvir2}, we find that the properties of the dominant formation channel can be different depending on the dynamical state of the clusters.
Moreover, we show that the ICL and BCG components in the central region come from the earlier infallers with a higher $M_{\rm{s}}/M_{\rm{h}}$ than ICL components in the outer region.
This can cause a radial gradient in the properties of the ICL and BCG progenitors to arise from the central region to the outer region in the cluster.

\begin{figure*}
\centering
\includegraphics[width=0.9\textwidth]{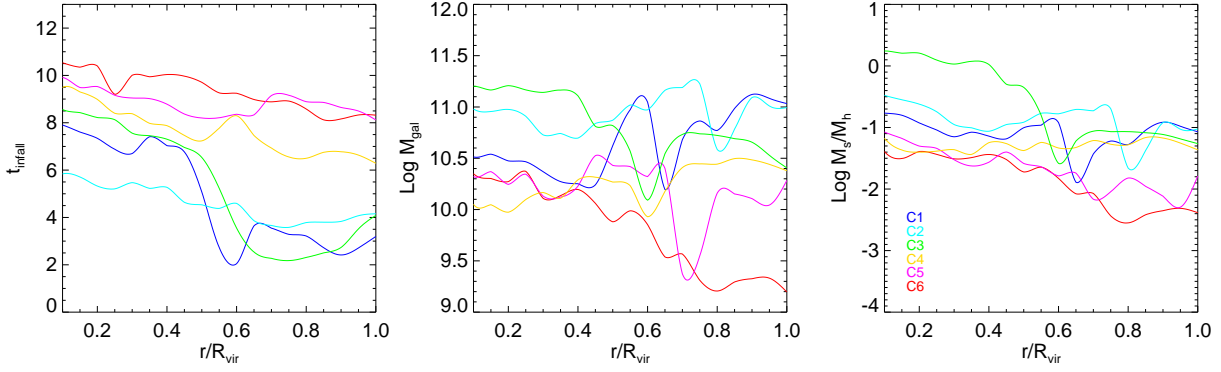}
\caption{Radial profile for the properties of progenitors of the ICL and BCG. From left to right, each panel indicates the radial \tcm{trend in the infall time ($t_{\rm{infall}}$), galaxy stellar mass (M$_{gal}$), and the total mass ratio of the progenitor to the cluster ($M_{\rm{s}}/M_{\rm{h}}$), respectively. Here $M_{\rm{gal}}$ and $M_{\rm{s}}/M_{\rm{h}}$ are the properties at $t_{\rm{infall}}$.} The blue, cyan, green, yellow, magenta, and red solid lines indicate C1-C6. The radius is normalized by $R_{\rm{vir}}$ of each cluster. \tcm{The lines in each panel are post-processed to show smoothly, but it does not affect the results.}}
\label{fig:profile}
\end{figure*}

Figure \ref{fig:profile} shows the radial profiles for the properties of the ICL and BCG in the six GRT clusters.
The radius is normalized by $R_{\rm{vir}}$ of each cluster, and the values of each property indicate the mean value of the progenitor's properties that contribute to the ICL and BCG within each 0.05$R_{\rm{vir}}$ interval from the center outwards.
The radial profiles of C1-C6 are colored blue, cyan, green, yellow, magenta, and red respectively.

The left panel of Figure \ref{fig:profile} shows a negative gradient with radius in the infall time of the progenitors for all clusters.
In particular, the gradient is steeper in C1 and C3.
This is because these clusters experience an ongoing merger with group-sized halos, but the halos have not yet reached the central region of the clusters by $z=0$.
Therefore, the stellar components of these `recent infallers' only contribute to the ICL components in the outer region.
Furthermore, we see another trend in which the progenitors contributing to the ICL and BCG of unrelaxed clusters fall into the clusters more recently than those of the relaxed clusters.
This trend is evident at all radii out to 1 $R_{\rm{vir}}$ of the clusters.

All clusters show a negative gradient in the infall time, whereas the radial trends in the stellar mass of the progenitors vary (the middle panel of Figure \ref{fig:profile}).
Nonetheless, we find that the ICL and BCG components of unrelaxed clusters form by more massive progenitors than those of relaxed clusters.
As we mentioned above, both the ICL and BCG in the unrelaxed clusters form by the progenitors falling into the cluster more recently than those of the relaxed cluster, and thus the progenitors have enough time to grow their mass by merging with other structures before their infall.
These hierarchical merging events naturally allow the presence of diffuse material such as a stellar halo in a massive galaxy or IGL in a group-sized halo before falling into the cluster \tcm{\citep[e.g.,][]{bell2008,darocha2005,darocha2008,spavone2018}}.
As the diffuse material can escape from the potential well of the host progenitor easier than the main stellar body, we can predict that the contribution of the massive galaxies to the ICL in the outer region of unrelaxed clusters comes from the diffuse material.
Indeed, in relaxed clusters, the `outer ICL' components are transferred to the cluster from their original host progenitors within 1.5-4~Gyr on average after falling into the cluster.
On the other hand, this transfer in the unrelaxed clusters occurs within 0.5-1.2~Gyr, and it is more rapid than in the relaxed clusters, as the progenitors include more diffuse material.
We will discuss the contribution from this diffuse material in Section \ref{sec:pre}.

The right panel of Figure \ref{fig:profile} generally shows a negative gradient with radius in the mass ratio of the progenitors (except for C4), similar to the $t_{\rm{infall}}$.
This negative gradient indicates that the progenitors with a higher mass ratio tend to reach the central region of the cluster by dynamical friction and are disrupted there.
However, the preferred mass ratio is different depending on the dynamical state of the clusters, i.e., the ICL and BCG components of the unrelaxed clusters are generated by progenitors with a higher mass ratio than the relaxed clusters at all radii in the cluster.

Even if $t_{\rm{infall}}$ and $M_{\rm{s}}/M_{\rm{h}}$ of all clusters show negative gradients, we conclude that their gradients are not similar because the range of slopes of the gradients is quite large, [-8,-1] for $t_{\rm{infall}}$ and [-2,-0.05] for $M_{\rm{s}}/M_{\rm{h}}$.
Our results also show that the variations in the trends are much stronger for the unrelaxed clusters because of their rapid recent mass growth.

\section{The effect of pre-processing on the formation of ICL and BCG} %%%%%%%%%%%%%%%%pre-processing
\label{sec:pre}

\begin{figure*}
\centering
\includegraphics[width=0.9\textwidth]{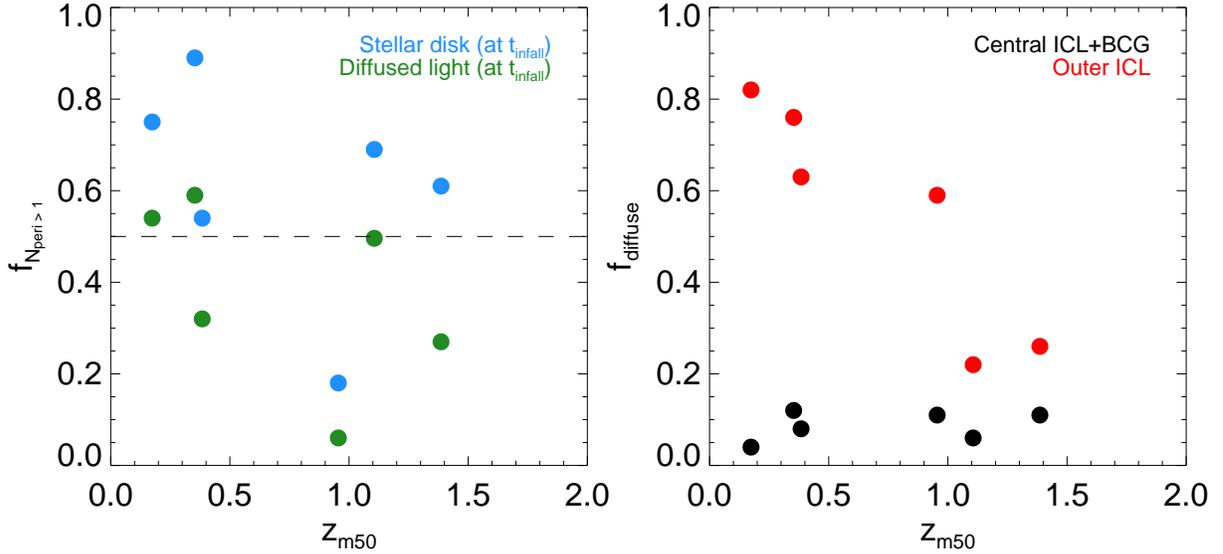}
\caption{\tcm{Contribution of `diffused light' and `stellar disk' to the ICL and BCG depending on $z_{\rm{m50}}$ of clusters. \textit{Left panel}: the fraction of the stellar components classified as N$_{peri} > 1$ among the stellar components that contribute to the ICL and BCG, where N$_{peri}$ is the number of times the stars pass through their pericenter while still bound to the progenitor. The filled blue and green circles indicate the `stellar disk' and `diffused light', respectively. \textit{Right panel}: the fraction of the `diffused light' among the stellar components that contribute to the `central ICL+BCG' (the filled black circles) and `outer ICL' (the filled red circles).}}
\label{fig:pre}
\end{figure*}

As tidal stripping tends to preferentially affect the outer region of the progenitors in the potential well of the cluster, the pre-processed diffuse material is stripped from the progenitors earlier than the stellar disk of the progenitors \tcm{\citep[e.g.,][]{smith2016}}.
Therefore, when/where the diffuse material and stellar disk transfer to the ICL and BCG will differ, even if they were inside the same progenitor.

To confirm their relative contribution, we investigate the contribution of the `diffused light' and `stellar disk' of the progenitors within the central region ($R < 0.1R_{\rm{vir}}$) and outer region ($R > 0.5R_{\rm{vir}}$).
In this paper, the `diffused light' indicates the diffuse material which are less dense than $\rho_{thresh} = 10^{-4.5}~M_{\odot} ~\rm{pc}^{-3}$ at $t_{\rm{infall}}$.
This $\rho_{thresh}$ was chosen because it was found to effectively separate the stars into those that are bound or unbound to satellite galaxies, in \cite {chun2022}.
\tcm{Note that we only use this simple 3D density threshold to measure the tidal stripping of `diffused light' and `stellar disk' from infalling satellites. This is because we do not separate the diffuse material and central galaxy using ROCKSTAR in this work, as we do in the ICL and BCG.}

To investigate the timing when the stellar components in the progenitors transfer to the ICL and BCG, we separate the `diffused light' and `stellar disk' into three subgroups depending on N$_{peri}$, where N$_{peri}$ is the number of times the \tcm{stars} pass through the pericenter while still bound to the progenitor. 
Those stars that transfer to the ICL and BCG before the host progenitor passes the first pericenter are classified as N$_{peri} = 0$, and the stellar components that move to the ICL and BCG after the host progenitor passes multiple pericenters are classified as N$_{peri} > 1$.
\tcm{The detailed data on the contribution of the `diffused light' and `stellar disk' depending on N$_{peri}$ to the `central ICL+BCG' and `outer ICL' is shown in Table \ref{tab:peri}.}

\tcm{The left panel of Figure \ref{fig:pre} shows how f$_{\rm{N_{peri} > 1}}$ relates to $z_{\rm{m50}}$ of clusters, where f$_{\rm{N_{peri} > 1}}$ is the fraction of the stellar components classified as N$_{peri} > 1$ subsample among the stellar components that contribute to the ICL and BCG.
In this panel, we show the f$_{\rm{N_{peri} > 1}}$ of combined `central ICL+BCG' and `outer ICL' because we do not find any trend difference between `central ICL+BCG' and `outer ICL'.
Note that although f$_{\rm{N_{peri} > 1}}$ of `stellar disk' and `diffused light' is decreasing with an increase of $z_{\rm{m50}}$, this trend is negligibly weak.}

\tcm{This panel show that the `stellar disk' (the filled blue circles) seems to move to the ICL and BCG less easily than the `diffused light' (the filled green circles). 
This can be seen by the fact that the f$_{\rm{N_{peri} > 1}}$ of `stellar disk' is higher than that of `diffused light' in all of the clusters.
Furthermore, more than half of the `stellar disk' is classified as N$_{peri} > 1$ subsample in the five clusters, whereas only two clusters in the case of `diffused light'.} %%%%%%%%%% 
This is because the `stellar disk' is located in the central region of the host progenitor, and thus its stripping is delayed until the outer region is stripped \tcm{\citep[e.g.,][]{smith2016}}.
Indeed, we find that the time for transition from the `stellar disk' of a galaxy to the `central ICL+BCG' is, on average, 0.4-0.8~Gyr longer than the `diffused light'.

\tcm{The right panel of Figure \ref{fig:pre} shows how f$_{\rm{diffuse}}$ relates to $z_{\rm{m50}}$ of clusters, where f$_{\rm{diffuse}}$ is the fraction of `diffused light' among the stellar components that contribute to the `central ICL+BCG' (the filled black circles) and 'outer ICL' (the filled red circles). }
In this panel, we can see that the contribution of the `diffused light' is small in the central region of all clusters, whereas its relative contribution increases in the outer region.
We find that, even if the relative contribution of `diffused light' increases, the absolute amount of the `diffused light' contributing to the `outer ICL' is smaller than the amount at `central ICL+BCG' because the total mass of `outer ICL' is only 1-9\% of total ICL and BCG.
However, in the outer region, we note the contribution of `diffused light' is strongly related to the dynamical state of the clusters\tcm{, i.e., t}he total contribution of `diffused light' increases as the $z_{\rm{m50}}$ decreases.
As described in Section \ref{sec:rvir1}, the progenitors of the `outer ICL' in unrelaxed clusters are usually `recent infallers' with a higher stellar mass compared to those in relaxed clusters.
This difference enables the progenitors of the `outer ICL' in unrelaxed clusters to gather more `diffused light' by the time of their infall.
In the case of unrelaxed clusters, we find that 11-13\% of the total stellar components in the progenitors of the `outer ICL' are `diffused light' at $t_{\rm{infall}}$, whereas, in the case of relaxed clusters, it is only 6-8\%.

\tcm{As we mentioned above, because the stripping of the `stellar disk' is delayed until the `diffused light' is stripped, most of the `stellar disks' contribute to the mass growth of the inner region of all the clusters, even if their progenitors dominate the mass growth of the `outer ICL'.}
Indeed, we find that 95\% of the disrupted components of `stellar disks' are located within $R < 0.5R_{\rm{vir}}$ for all the clusters.

\begin{figure*}
\centering
\includegraphics[width=0.9\textwidth]{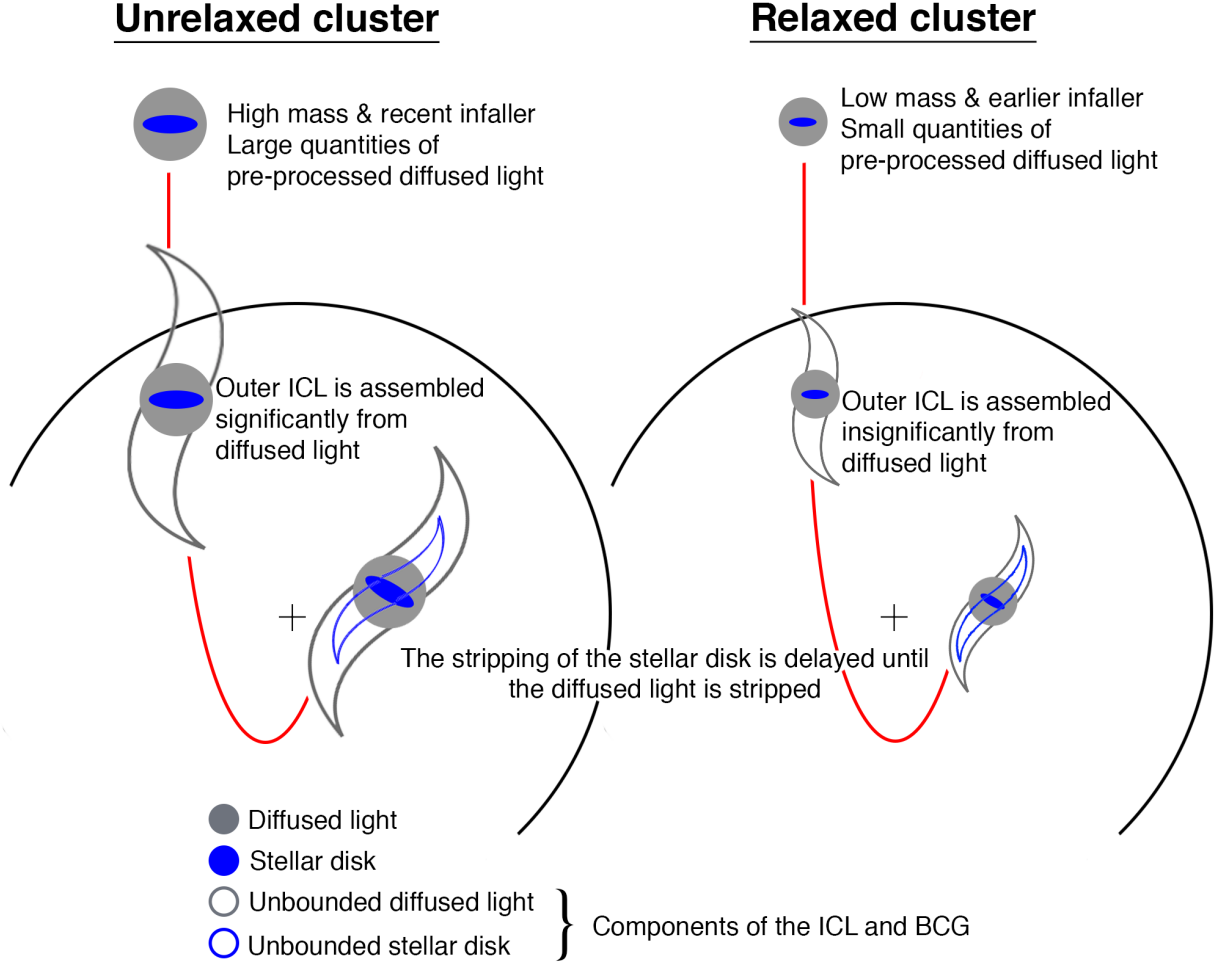}
\caption{Cartoon schematic of the mass loss of a progenitor of the ICL and BCG in the host cluster. The black arcs of the left and right panels indicate the $R_{\rm{vir}}$ of the unrelaxed and relaxed clusters. The gray and blue regions in the progenitor of the ICL and BCG represent the components of pre-processed `diffused light' and `stellar disk'. These components are tidally stripped from the host progenitor in the cluster and will contribute to the ICL and BCG.}
\label{fig:cartoon}
\end{figure*}

\tcm{Figure \ref{fig:cartoon} summarizes our main results using a cartoon schematic of the mass loss of an example progenitor of the ICL and BCG in the unrelaxed and relaxed clusters.}
The figure illustrates how the contribution of pre-processed `diffused light' to the `outer ICL' is more significant in the unrelaxed clusters than in the relaxed clusters, as more massive infallers include more `diffused light'. 
Furthermore, we can see that the stripping of the `stellar disk' is delayed until the `diffused light' is stripped, regardless of the dynamical state of the host clusters.

\section{Discussion}
\label{sec:discussion}
(1) The progenitors of the ICL and BCG. - In Section \ref{sec:rvir1}, we show that the contribution of `recent infallers' ($z_{\rm{infall}} < 1$) decreases as the clusters are more relaxed.
In particular, their contribution is less than 50\% in the two most relaxed clusters.
Despite this fact, we find that most of the progenitors of the `outer ICL' in all clusters are stripped from their host galaxy during the last 8~Gyr ($z<1$) and then belong to the `outer ICL'.

In the central region ($R < 0.1R_{\rm{vir}}$), the contribution of progenitors that fall into the cluster earlier than 8~Gyr ago is more significant for all clusters.
Nonetheless, we find that over half of the `central ICL+BCG' in all clusters except for C6 are stripped from their host galaxy during the last 8~Gyr. This dominant mass growth of the ICL and BCG during the last 8~Gyr is in agreement with the previous observation and simulation studies \citep{murante2007,montes2014,contini2018}.

Unlike in other clusters, 73\% of the `central ICL+BCG' components in the most relaxed cluster, C6, were already generated before 8~Gyr.
Although the BCG is dominant in the central region, this indicates that a significant fraction of the ICL components in the central region can be generated prior to $z\sim1$, as suggested by \cite{ko2018}.

As we mentioned in Section \ref{sec:infall}, the progenitors of the ICL and BCG are generally more massive than $M_{\rm{gal}} \sim 10^{10}~M_{\odot}~h^{-1}$.
In particular, we find that half of the ICL and BCG components in $R_{\rm{vir}}$ of the unrelaxed clusters come from the massive galaxies and/or group-sized halos with $M_{\rm{gal}} > 10^{11}~M_{\odot}$. 
This contribution from massive progenitors corresponds with the results of \cite{pillepich2018}. 
On the other hand, in the case of the relaxed clusters, half of the ICL and BCG stars are from smaller progenitors than \tcm{typical} progenitors in the unrelaxed clusters, but they are still composed of massive galaxies with $M_{\rm{gal}} > 10^{10.5}~M_{\odot}$.
Although we include the BCG, our results confirm the results of previous studies that suggest the massive galaxies are the dominant formation channel of the ICL \tcm{\citep[e.g.,][]{contini2014,montes2018}}.

(2) Evolution of ICL and BCG fraction depending on the dynamical state of the clusters. - At $z=0$, f$_{\rm{ICL+BCG}}$ of clusters increases with $z_{\rm{m50}}$ of the clusters (the right panel of Figure \ref{fig:relation}). 
The fact that the clusters have a unique mass growth history implies that f$_{\rm{ICL+BCG}}$ of clusters also evolves differently from each other.
We find that f$_{\rm{ICL+BCG}}$ of relaxed clusters increases 13-24 \tcm{percentage points} from $z=1$ until $z=0$ by continuous disruption of satellites.
On the other hand, f$_{\rm{ICL+BCG}}$ of unrelaxed clusters is significantly affected by the infall of massive group-sized halos after $z=1$.
As the infall of group-sized halos increases the total stellar mass in unrelaxed clusters, f$_{\rm{ICL+BCG}}$ decreases.
However, because their tidal disruption eventually contributes to the mass growth of the ICL and BCG, the degree of tidal disruption regulates f$_{\rm{ICL+BCG}}$.
Indeed, we find that f$_{\rm{ICL+BCG}}$ of unrelaxed clusters increases slightly compared to that of relaxed clusters or even decreases after $z=1$.

\cite{montes2022} showed no significant evolution of f$_{\rm{ICL+BCG}}$ in $0<z<0.6$ using the observational data from \cite{gonzalez2007} and \cite{furnell2021}.
Our results show the possibility that this insignificant time evolution of f$_{\rm{ICL+BCG}}$ can result from the different dynamical states of the clusters.
Therefore, to enhance our understanding of the evolution of f$_{\rm{ICL+BCG}}$, we believe it will be necessary for the future to investigate larger samples of simulated clusters with various dynamical states.

(3) The pre-processed `diffused light' of surviving satellites. - as mentioned in Section \ref{sec:GRT}, we do not include the stellar halo of massive galaxies and IGL of group-sized halos in the host cluster to the ICL and BCG components because we define the ICL as the stellar components that are bound to only the cluster.
This choice of whether we include them in the ICL can change the total amount of the ICL and BCG significantly.
Indeed, we find that f$_{\rm{ICL+BCG}}$ of all clusters increases if we add the `diffused light' defined as the stellar components, which are less dense than $\rho_{\rm{thresh}}$ to the ICL components.
The increasing fraction is related to the dynamical state of clusters. 
For example, the fraction of ICL and BCG in the most relaxed cluster, C6, increases by \tcm{3 percentage points}, but that in C1 increases by \tcm{10 percentage points} as the cluster experiences the ongoing merger with two group-sized halos, which have IGL components.

In Section \ref{sec:rvir1}, we mentioned that the amount of `outer ICL' in C2 is 9\% of the total stellar mass in the cluster.
This fraction is very high compared to other clusters that have only 1-2\%.
We find that half of the `outer ICL' in C2 was stripped from the IGL components of three group-sized halos falling into the cluster after $z\sim0.4$.
In addition, if the IGL components of surviving group-sized halos are stripped from the host halos and added to the ICL and BCG, the amount of the ICL and BCG can be increased to 30\% of the total stellar mass in the cluster.
This increase demonstrates the consequences of the pre-processing of galaxies in the group environment, as shown by the diffuse material near NGC 4365 in the Virgo cluster \citep{mihos2017}.

\cite{rudick2006,rudick2009,rudick2011} investigated the ICL using the cosmological N-body simulations with a resimulation technique similar to the GRT.
However, as the authors replaced the original DM halos with the higher-resolution galaxies at $z=2$, the resimulation only begins at $z=2$.

In Section \ref{sec:rvir2}, we show that the contribution of `early infallers' ($z_{\rm{infall}} > 1.7$) to `central ICL+BCG' is insignificant, except for C6.
Nevertheless, replacing the original DM halos with the high-resolution galaxies at $z=2$ can lead to unrealistic evolution of the clusters, as the halos may have already experienced mass loss tidally within the more massive halo before $z=2$.
On the other hand, since the GRT simulations start their resimulation before $z=6.5$ and replace the halos with the high-resolution galaxies when they have $M_{\rm{peak}}$, the halos do not experience such mass loss. 
Therefore, we can trace the evolution of clusters more realistically than previous similar works \citep{rudick2006,rudick2009,rudick2011}.
These advantages are useful for comparison with the observation beyond $z=1$ \citep{ko2018} and for studying the evolution of the ICL and BCG at high redshift.

\section{Summary}
\label{sec:summary}
Using cosmological N-body simulations, we investigate the formation channels of the ICL and BCG in the six clusters of $M_{\rm{vir}} \sim 1-2\times10^{14}~M_{\odot}~h^{-1}$.
To describe the evolution of the stellar structures in cosmological N-body simulations, we adopt an alternative simulation technique referred to as the ``Galaxy Replacement Technique" (GRT), first introduced in \cite{chun2022}.
GRT does not include computationally expensive hydrodynamic recipes.
Instead, we replace low-resolution dark matter halos with high-resolution models of galaxies, including a dark matter halo and stellar disk.
This inexpensive simulation technique with high-resolution particles enables us to accurately resolve the tidal stripping process and model diffuse features with sufficient star particles.

In this study, we wish to understand how the formation channels of the ICL and BCG differ depending on the dynamical state of the clusters, and as a function of distance from the cluster centers. We also study how significant pre-processing is for the growth of the ICL and BCG.
Therefore, we investigate the progenitors of the `central ICL+BCG' ($R < 0.1R_{\rm{vir}}$) and `outer ICL' ($R > 0.5R_{\rm{vir}}$) in the six clusters, which evolve on different time scales, and trace their properties at t$_{\rm{infall}}$.
We summarize our results as follows:

\begin{enumerate}

\item[(1)] The fraction of the ICL and BCG, f$_{\rm{ICL+BCG}}$, increases with $z_{\rm{m50}}$, which represents the dynamical state of the clusters \tcm{(the right panel of Figure \ref{fig:relation})}. \tcm{Here, the $z_{\rm{m50}}$ is defined as the epoch when the cluster first acquires half of its virial mass at $z=0$.}

\item[(2)] \tcm{As the cluster is more dynamically unrelaxed, the contribution of the `recent infallers' ($t_{\rm{infall}} < 8~$Gyr) to the `outer ICL' increases.}
This results in the infalling progenitors being more massive, as they have more time to grow their mass outer the cluster.
However, the \tcm{typical} progenitors of the `outer ICL' fall into all clusters with \tcm{`low-ratio infallers'}, $M_{\rm{s}}/M_{\rm{h}} < 0.1$, regardless of the dynamical state of the clusters \tcm{(see Table \ref{tab:radius05} and Figure \ref{fig:cont05})}. \tcm{Here $M_{\rm{s}}/M_{\rm{h}}$ is defined as the total mass ratio of the ICL progenitor to the cluster at $t_{\rm{infall}}$.}

\item[(3)] \tcm{As the cluster is more dynamically relaxed, the contribution of `early infallers' ($t_{\rm{infall}} > 10~$Gyr) to the `central ICL+BCG' increases.}
The \tcm{typical} progenitors of the `central ICL+BCG' in all clusters are generally massive halos of $M_{\rm{gal}} > 10^{10}~M_{\odot}~h^{-1}$ and $M_{\rm{s}}/M_{\rm{h}} > 0.1$ \tcm{(see Table \ref{tab:radius01} and Figure \ref{fig:cont01})}.

\item[(4)] At all radii, the progenitors of the ICL and BCG in the relaxed clusters fall in earlier, with lower $M_{\rm{gal}}$, and lower $M_{\rm{s}}/M_{\rm{h}}$, than those in unrelaxed clusters.
Furthermore, we find negative radial gradients in $t_{\rm{infall}}$ and $M_{\rm{s}}/M_{\rm{h}}$ of the progenitors of the ICL and BCG in all our clusters \tcm{(see Figure \ref{fig:profile})}.

\item[(5)] The contribution of pre-processed `diffused light' to the `outer ICL' increases when the cluster is less relaxed because massive progenitors with more abundant IGL frequently fall into unrelaxed clusters \tcm{(see Figure \ref{fig:pre} and Table \ref{tab:peri})}.
\end{enumerate}

Our results show that the formation channel of the ICL and BCG can be different depending on the dynamical state and distance from the center of the clusters.

In this paper, we focus on clusters with similar masses.
However, in the future, we will extend our study to investigate the formation channel of the diffuse light and central galaxy of systems with a greater range of mass, including less massive group-sized halos and more massive clusters, and with a broad range in dynamical state.

%We are grateful to the anonymous referee for the helpful comments.
%KWC was supported by the National Research Foundation of Korea(NRF) grant funded by the Korea government(MSIT) (2021R1F1A1045622).
%JHS acknowledges support from the National Research Foundation of Korea grant (2021R1C1C1003785) funded by the Ministry of Science, ICT \& Future Planning.
%JWY was supported by a KIAS Individual Grant (QP089901) via the Quantum Universe Center at Korea Institute for Advanced Study.
%This work was also supported by the National Supercomputing Center with supercomputing resources including technical support (KSC-2020-CRE-0297).

\appendix
\section{Tabulated data on the effect of pre-processing}
\label{sec:appendix}
\restartappendixnumbering

 \begin{deluxetable*}{ccccc|c|cccc|c|}[b]
%\tablenum{1}
\tablecaption{Effect of pre-processing}
%\tablewidth{0pt}
\tablehead{
\colhead{} & \colhead{} & \multicolumn4c{Diffused light (at $t_{\rm{infall}}$)} & \colhead{} & \multicolumn4c{Stellar disk (at $t_{\rm{infall}}$)} \\
\cline{3-6} \cline{8-11}
\colhead{} & \colhead{} & \colhead{N$_{peri} = 0$} & \colhead{N$_{peri} = 1$} & \colhead{N$_{peri} > 1$} & \colhead{Total} & \colhead{} & \colhead{N$_{peri} = 0$} & \colhead{N$_{peri} = 1$} & \colhead{N$_{peri} > 1$} & \colhead{Total}
}
\startdata
\multirow{6}{1em}{\rotatebox{90}{Central ICL+BCG}} & C1 & 0.00 & 0.01 & \textbf{0.03} & 0.04 & & 0.15 & 0.10 & \textbf{0.71} & 0.96\\
& C2 & 0.01 & 0.03 & \textbf{0.08} & 0.12 & & 0.06 &  0.03 & \textbf{0.79} & 0.88\\
& C3 & \textbf{0.04} & 0.01 & 0.03 & 0.08 & & 0.42 & 0.01 & \textbf{0.49} & 0.92\\
& C4 & 0.03 & \textbf{0.08} & 0.00 & 0.11 & & 0.06 & \textbf{0.67} & 0.16 & 0.89\\
& C5 & 0.00 & \textbf{0.03} & 0.03 & 0.06 & & 0.08 & 0.22 & \textbf{0.64} & 0.94\\
& C6 & 0.01 & \textbf{0.07} & 0.03 & 0.11 & & 0.10 & 0.25 & \textbf{0.54} & 0.89\\
\hline
\multirow{6}{1em}{\rotatebox{90}{Outer ICL}} & C1 & \textbf{0.40} & 0.19 & 0.23 & 0.82 & & 0.01 & 0.02 & \textbf{0.15} & 0.18\\
& C2 & 0.12 & 0.25 & \textbf{0.39} & 0.76 & & 0.03 & 0.01 & \textbf{0.20} & 0.24\\
& C3 & 0.17 & \textbf{0.30} & 0.16 & 0.63 & & 0.09 & 0.02 & \textbf{0.26} & 0.37\\
& C4 & 0.03 & \textbf{0.33} & 0.23 & 0.59 & & 0.09 & 0.13 & \textbf{0.19} & 0.41\\
& C5 & 0.02 & 0.05 & \textbf{0.15} & 0.23 & & 0.11 & 0.03 & \textbf{0.63} & 0.77\\
& C6 & 0.03 & \textbf{0.18} & 0.05 & 0.26 & & 0.00 & 0.22 & \textbf{0.52} & 0.74\\
\enddata
\tablecomments{The contribution of the stellar components to the ICL and BCG depending on the number of times the \tcm{stars} pass through the pericenter while still bound to the progenitor, N$_{peri}$. The stellar components of the progenitors are categorized as `diffused light' and `stellar disk' by their density at $t_{\rm{infall}}$. The threshold between the `diffused light' and `stellar disk' is $\rho_{thresh} = 10^{-4.5}~M_{\odot} ~\rm{pc}^{-3}$. Column 5 and 9 indicate the total fraction of `diffused light' and `stellar disk', respectively. The first six rows are for the `central ICL+BCG' and the second six rows are for `outer ICL'. The dominant formation channel is indicated in bold.} 
\label{tab:peri}
%\vspace{-1cm}
\end{deluxetable*}

\bibliography{main}{}
\bibliographystyle{aasjournal}

\end{document}